\documentclass[english,11pt,draftcls,onecolumn]{IEEEtran}
\usepackage[T1]{fontenc}
\usepackage[latin9]{inputenc}
\usepackage{color}
\usepackage{babel}
\usepackage{float}
\usepackage{amsthm}
\usepackage{amsmath}
\usepackage{amssymb}
\usepackage{graphicx}
\usepackage[unicode=true,
 bookmarks=true,bookmarksnumbered=true,bookmarksopen=true,bookmarksopenlevel=1,
 breaklinks=false,pdfborder={0 0 0},backref=false,colorlinks=false]
 {hyperref}
\hypersetup{pdftitle={Your Title},
 pdfauthor={Your Name},
 pdfpagelayout=OneColumn,pdfnewwindow=true,pdfstartview=XYZ,plainpages=false}

\makeatletter

\floatstyle{ruled}
\newfloat{algorithm}{tbp}{loa}
\providecommand{\algorithmname}{Algorithm}
\floatname{algorithm}{\protect\algorithmname}

\theoremstyle{plain}
\newtheorem{thm}{\protect\theoremname}
\theoremstyle{plain}
\newtheorem{lem}[thm]{\protect\lemmaname}

\@ifundefined{definecolor}
 {\usepackage{color}}{}
\usepackage{babel}
\@ifundefined{definecolor}{\usepackage{color}}{}
\usepackage{babel}
\ifCLASSOPTIONcompsoc
\else
\fi

\makeatother

\providecommand{\lemmaname}{Lemma}
\providecommand{\theoremname}{Theorem}

\begin{document}

\title{Optimizing Relay Precoding for Wireless Coordinated Relaying}

\author{\authorblockN{Lilian Hong} \authorblockA{\\
Lawrence Berkeley National Laboratory\\
 Berkeley, CA}}

\maketitle

\section{Introduction }

Recently there have been extensive studies on cooperative, relay-based
transmissions for extending cellular coverage or increasing diversity.
Several basic relaying techniques have been introduced, such as amplify-and-forward
(AF) \cite{Farhadi:AF,fan:overhearing}, decode-and-forward \cite{Zhu:DF,chan:cdr2}
and compress-and-forward \cite{Uppal:CF}.

These transmission techniques have been applied in one-way, two-way
or multi-way relaying scenarios. There has been a particularly high
interest in two-way relaying scenarios \cite{Petar:two_way,Katabi:ANC,fan:spawc,Ning:overhearing,Huaping:completion},
where throughput gains have been demonstrated by utilizing the ideas
of wireless network coding \cite{Huaping:4WR,Huaping:4WR2}. The two
underlying principles used in designing throughput\textendash{}effi{}cient
schemes with wireless network coding:
\begin{enumerate}
\item Aggregation of multiple communication fl{}ows: instead of transmitting
each fl{}ow independently, network coding is used where fl{}ows are
sent/processed jointly;
\item Network coding intentionally allows interference and simultaneous
usage of the shared wireless medium, leaving to the receivers to remove
the adverse impact of interference by using any side information.
\end{enumerate}
Leveraging on these principles, there are proposed schemes with AF
relaying in \cite{fan:CDR2,Fan:CDR1} that feature more general traffic
patterns compared to the two-way relaying. These schemes are termed
\emph{coordinated direct/relay} (CDR) transmissions. The CDR transmission
considers scenarios where one direct user (UE) and one relayed UE
are served in uplink/downlink. The relayed UE is assumed to have no
direct link to the base station (BS) due to large path loss and relies
only on the amplified/forwarded signal from the relay in order to
decode the signal from the BS. Schemes that are related to some of
the schemes have appeared before in \cite{Paulraj:overhearing,Yomo:IC,Letaief:IC}.

Each user might have a downlink or uplink traffic. Hence, there are
different traffic configurations. We focus on one representative traffic
type with one relayed uplink UE and one direct downlink UE. This case
displays the merits of analog network coding in a setting that is
more general than the usual two-way relay scenario. Furthermore it
showcases the principle of overheard information where a node overhears
a signal that is not intended to itself and uses it as \emph{a priori}
information to cancel interference in an ulterior transmission phase.

\begin{figure}[tbh]
\vspace{-15pt}

\centering \includegraphics[scale=0.6]{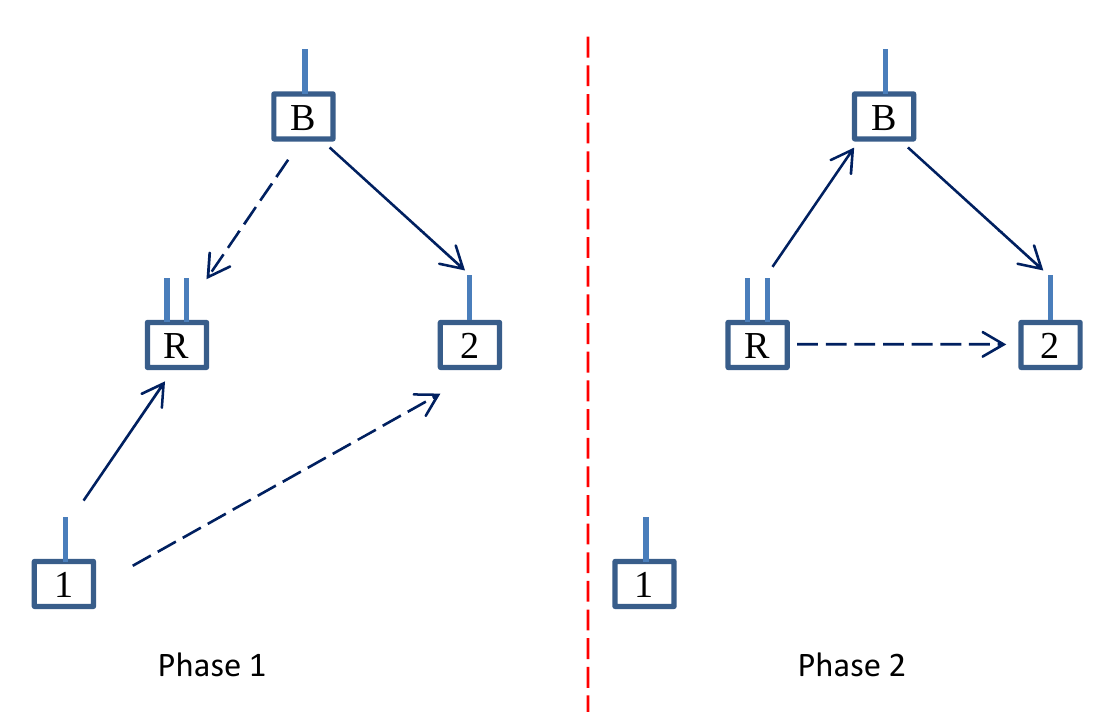}\caption{CDR MIMO System Model.}

\vspace{-15pt}

\label{Flo:fig_CDR_MIMO_DL_Model}
\end{figure}

In the scheme on Fig. \ref{Flo:fig_CDR_MIMO_DL_Model}, we assume
that a relayed UE has one signal to deliver to the BS through the
assistance of the relay station, while a direct user wants to receive
a signal from the BS. Notice in a conventional wireless cellular system,
these signals are sent over two orthogonal uplink and downlink phases
for the two separate information flows, respectively. Instead in the
CDR system, the BS first sends the signal to the direct UE and simultaneously
the relayed UE transmits the signal to the relay station in phase
1. The relay receives two signals: the desired signal from the UE
and an interfering signal from the BS. It does not decode the signals
but instead forward them in phase 2 using the principle of analog
network coding. The simultaneous two-flow transmissions improve the
\emph{spectral efficiency} compared to the conventional method. The
key points are the BS can use the \emph{a priori }information to perform
self-interference cancellation and enable interference-free reception
and decoding; the direct UE can use the overheard information in phase
2 to help decoding the desired signal.

In the works that deal with the CDR transmission, the relay has a
central role in managing the interference. Therefore, in this work
we investigate the qualitative changes and the performance improvements
that arise when the relay node in the CDR schemes is equipped with
multiple antennas. Differently from the previous works, the usage
of multiple antennas at the relay permits to manage the interference
and boost the overall system performance through beamforming. This
is a significant conceptual difference compared to the original CDR
schemes, while the usage of multiple antennas at the BS and the UEs
is a clear future work. We consider AF operation at the relay, assuming
that the relay and the reception nodes have a perfect channel state
information (CSI). Our objective is to maximize the achievable sum-rate
of the system. Our design shows that the overall system performance
is improved by allowing the relay beamformer to deliver the interfered
signal to both the BS and the direct UE in phase 2. Meanwhile, the
BS completely cancels the self-interference; the direct UE applies
linear interference minimization receiver to decode the desired signal.

We propose three low-complexity algorithms to approach an upper bound
on the sum-rate, namely the adaptive subspace averaging algorithm,
the power iteration algorithm and the linear space spanning algorithm.
Their performance is shown via simulations to be close to the tight
upper bound on the sum-rate. The gain via possessing multiple relay
antennas is also shown compared to the original CDR transmission.

\emph{Notation}: We use uppercase and lowercase boldface letters to
represent matrices and vectors, respectively. $\otimes$ refers to
the Kronecker product and $||\mathbf{\cdot}||_{F}^{2}$ denotes the
Frobenius norm of a matrix. $\mathbb{I}$ is the identity matrix.

\section{System Overview\label{Section 2} }

The basic setup is the scenario in Fig. \ref{Flo:fig_CDR_MIMO_DL_Model}
with one BS, one relay, and two UEs. The relay is equipped with $M$
antennas. The BS and the UEs are equipped with one antenna only. The
transmission from the relayed UE to the relay has the same duration
as the transmission from the relay to the BS. The relay is deployed
to help the relayed UE which has no direct link to the BS due to large
path loss.

We consider the multi-antenna relay beamforming design where there
are two information flows: the relayed UE (UE 1) delivers $x_{1}$
to the BS and the BS transmits $x_{2}$ to the direct UE (UE 2). The
conventional system will create two orthogonal transmissions for separate
information flows, while the CDR system enables simultaneous transmissions
and thus improves the system spectral efficiency. Here we illustrate
the two-phase CDR transmission in Fig. \ref{Flo:fig_CDR_MIMO_DL_Model}.
In the first slot, UE 1 transmits $x_{1}$ to the relay and the BS
delivers $x_{2}$ to UE 2 simultaneously. At the same time, UE 2 overhears
the signal from UE 1 and the relay also receives the signal from the
BS. Then the relay forwards the received \textcolor{black}{physical
layer network-coded} signal to both the BS and the direct UE in the
second slot. The BS has the capability to use the \emph{a priori}
information it transmits in the first slot to perform self-interference
cancellation. In summary, this gives
a more general traffic pattern compared to the two-way relaying.

In this CDR system, each channel is assumed to be an independent complex
Gaussian random variable with zero mean and unit variance. All links
are assumed to be static within the two slots. Assume $P$ to be the
transmit power of the BS and each UE, the received signals at the
relay and UE 2 in the first slot are

%\begin{figure}[tbh]
%\vspace{-11pt}
%
%\centering \includegraphics[scale=0.6]{Two_Flows}\caption{Two Information Flows in CDR.}
%
%
%\vspace{-15pt}
%
%\label{Flo:fig_CDR_two_flows}
%\end{figure}

\vspace{-10pt}

\begin{eqnarray}
\mathbf{y}_{R} & = & \sqrt{P}\mathbf{h}_{R1}x_{1}+\sqrt{P}\mathbf{h}_{RB}x_{2}+\mathbf{n}_{R}\nonumber \\
y_{2}[1] & = & \sqrt{P}h_{21}x_{1}+\sqrt{P}h_{2B}x_{2}+n_{2}[1]\label{eq:signal_first_slot}
\end{eqnarray}
where $\mathbf{n}_{R}$ is the complex white Gaussian noise vector
at the relay with the covariance matrix $\mathbb{E}[\mathbf{n}_{R}\mathbf{n}_{R}^{{\rm H}}]=\mathbb{I}$
and $n_{2}[1]$ is the complex white Gaussian noise variable at UE
2 in the first slot with unit variance%
\footnote{We assume the variance of each noise component is normalized.%
}. The received signals at the BS and UE 2 in the second slot are

\vspace{-20pt}

\begin{eqnarray}
y_{B} & = & \mathbf{h}_{BR}\mathbf{\mathbf{x}}_{R}+n_{B}\nonumber \\
y_{2}[2] & = & \mathbf{h}_{2R}\mathbf{x}_{R}+n_{2}[2]\label{eq:signal_second_slot}
\end{eqnarray}
where the signal vectors transmitted from the relay is in the form
$\mathbf{\mathbf{x}}_{R}=\mathbf{W}\mathbf{\mathbf{y}}_{R}$ with
$\mathbf{W}$ being the $M\times M$ relay beamforming matrix. At
the relay, $\mathbf{W}$ is used here to linearly process $M\times1$
received signal vector and form the $M\times1$ transmit signal vector
without loss of generality. $n_{B}$ and $n_{2}[2]$ are the complex
white Gaussian noise variables with unit variance each at the BS and
UE 2 respectively. The total relay power is constrained not to exceed
a power budget $\mathbb{E}[\mathbf{x}_{R}^{{\rm H}}\mathbf{\mathbf{x}}_{R}]=P(\mathbf{h}_{RB}^{{\rm H}}\mathbf{W}^{{\rm H}}\mathbf{W}\mathbf{h}_{RB}+\mathbf{h}_{R1}^{{\rm H}}\mathbf{W}^{{\rm H}}\mathbf{W}\mathbf{h}_{R1})+||\mathbf{W}||_{F}^{2}\leq P_{R}$.

\section{Achievable Sum-Rate Maximization for Coordinated Relay Beamforming}

From the previous illustration, the relay has the capability to beamform
the received network-coded signal and forwards the beamformed signal
to both the BS and the direct UE in the second phase of the CDR transmission.
Via sum-rate maximal relay beamforming design, the overall CDR system
performance is enhanced by allowing the relay to balance between maximizing
the rate of the transmission from the relayed UE to the BS and rate
of the transmission from the BS to the direct UE. The central role
of the relay in balancing the two information flows can be observed
in Fig. \ref{Flo:fig_CDR_two_flows}. In this section, we focus on
the problem of achievable sum-rate maximization subjecting to the
total relay power constraint. Using the relay beamforming matrix as
the design parameter, the problem is shown to be equivalent to maximizing
the product of two fractional quadratic functions. A tight upper performance
bound on the sum-rate will be given first and three low-complexity
solutions will be provided to approach the optimal solution of the
non-convex problem.

\subsection{Problem Formulation}

The sum-rate maximization problem can be formulated as

\begin{align*}
\arg\max_{\mathbf{W}}\:\: & \left(R_{1}+R_{2}\right)\\
s.t. & \;\: P(\mathbf{h}_{RB}^{{\rm H}}\mathbf{W}^{{\rm H}}\mathbf{W}\mathbf{h}_{RB}+\mathbf{h}_{R1}^{{\rm H}}\mathbf{W}^{{\rm H}}\mathbf{W}\mathbf{h}_{R1})\\
 & \;\:+||\mathbf{W}||_{F}^{2}\leq P_{R}
\end{align*}
where $R_{1}$ and $R_{2}$ denote the rate expressions for the transmission
of $x_{1}$ and $x_{2}$, respectively. The rate expression for each
information flow can be written as $R_{1}=\frac{1}{2}\log_{2}(1+{\rm SNR_{1}})$
and $R_{2}=\frac{1}{2}\log_{2}(1+{\rm SINR_{2}})$ where ${\rm SNR_{1}}$
is the SNR expression for the BS to decode $x_{1}$ and ${\rm SINR_{2}}$
is the SINR expression for the direct UE to decode $x_{2}$. And the
factor $\frac{1}{2}$ is due to the two time slots transmission duration.
This is because from the analog network coding principle, $x_{2}$
is known \emph{a priori} at the BS and the related interference is
mitigated via the self-interference cancellation process. Therefore,
there is no interference when the BS wants to decode $x_{1}$. Notice
we then use linear receivers in the CDR system to decode the desirable
signals at the BS and the direct UE. Using the monotonicity of the
$\log$ function, the sum-rate maximization problem can be rewritten
as

\begin{align}
\arg\max_{\mathbf{W}}\:\: & \left[(1+{\rm SNR_{1}})(1+{\rm SINR_{2}})\right]\nonumber \\
s.t. & \;\: P(\mathbf{h}_{RB}^{{\rm H}}\mathbf{W}^{{\rm H}}\mathbf{W}\mathbf{h}_{RB}+\mathbf{h}_{R1}^{{\rm H}}\mathbf{W}^{{\rm H}}\mathbf{W}\mathbf{h}_{R1})\label{eq:sum_rate_max_criterion}\\
 & \:\;+||\mathbf{W}||_{F}^{2}\leq P_{R}.\nonumber
\end{align}

We first take a look at the SNR and SINR expressions for both UEs.
For the BS, after self-interference cancellation, we will have $\hat{y}_{B}=\sqrt{P}\mathbf{h}_{BR}\mathbf{W}\mathbf{h}_{R1}x_{1}+\mathbf{h}_{BR}\mathbf{W}\mathbf{n}_{R}+n_{B}$.
Then the SNR at the BS is expressed as
\[
{\rm SNR_{1}=}\frac{P\mathbf{h}_{BR}\mathbf{W}\mathbf{h}_{R1}\mathbf{h}_{R1}^{{\rm H}}\mathbf{W}^{{\rm H}}\mathbf{h}_{BR}^{{\rm H}}}{\mathbf{h}_{BR}\mathbf{W}\mathbf{W}^{{\rm H}}\mathbf{h}_{BR}^{{\rm H}}+1}.
\]

Meanwhile, the direct UE uses $y_{2}[1]$ from the first slot and
$y_{2}[2]$ from the second slot to from a virtual 2-antenna received
signal vector $\mathbf{y}_{2}=\left[\begin{array}{cc}
y_{2}[1] & y_{2}[2]\end{array}\right]^{{\rm T}}$

\begin{align*}
\mathbf{y}_{2} & =\left[\begin{array}{c}
\sqrt{P}h_{2B}\\
\sqrt{P}\mathbf{h}_{2R}\mathbf{W}\mathbf{h}_{RB}
\end{array}\right]x_{2}+\left[\begin{array}{c}
\sqrt{P}h_{21}\\
\sqrt{P}\mathbf{h}_{2R}\mathbf{W}\mathbf{h}_{R1}
\end{array}\right]x_{1}\\
 & \quad\;+\left[\begin{array}{c}
n_{2}[1]\\
\mathbf{h}_{2R}\mathbf{W}\mathbf{n}_{R}+n_{2}[2]
\end{array}\right].
\end{align*}
Then the direct UE wants to estimate the desired signal $x_{2}$ and
$x_{1}$ is the interference from the other information flow. We use
simple zero forcing (ZF) receiver at the direct UE to aim for a low
computational complexity \cite{Elisabeth:book,Fan:CDR3}. The corresponding
SINR at UE 2 is derived as

\[
{\rm SINR_{2}=}\frac{P\left\Vert h_{2B}\mathbf{h}_{2R}\mathbf{W}\mathbf{h}_{R1}-h_{21}\mathbf{h}_{2R}\mathbf{W}\mathbf{h}_{RB}\right\Vert _{2}^{2}}{|h_{21}|^{2}\left(\mathbf{h}_{2R}\mathbf{W}\mathbf{W}^{{\rm H}}\mathbf{h}_{2R}^{{\rm H}}+1\right)+\left\Vert \mathbf{h}_{2R}\mathbf{W}\mathbf{h}_{R1}\right\Vert _{2}^{2}}.
\]

The following lemma summarizes the main result of the problem formulation
and is proved in the Appendix.
\begin{lem}
The sum-rate maximization beamforming design is equivalent to maximizing
the product of two fractional quadratic functions
\begin{equation}
\arg\max_{\tilde{\mathbf{w}}}G(\tilde{\mathbf{w}})=\arg\max_{\tilde{\mathbf{w}}}\left[\frac{\tilde{\mathbf{w}}^{{\rm H}}\mathbf{A}\tilde{\mathbf{w}}}{\mathbf{\tilde{w}}^{{\rm H}}\mathbf{B}\mathbf{\tilde{w}}}\times\frac{\tilde{\mathbf{w}}^{{\rm H}}\mathbf{C}\tilde{\mathbf{w}}}{\mathbf{\tilde{w}}^{{\rm H}}\mathbf{D}\mathbf{\tilde{w}}}\right]\label{eq:lemma-1}
\end{equation}
where matrices $\mathbf{A}$, $\mathbf{B}$, $\mathbf{C}$, and $\mathbf{D}$
are not dependent on $\tilde{\mathbf{w}}$. Then $\mathbf{\tilde{w}}$
is scaled to fulfill the power constraint $\tilde{\mathbf{w}}^{{\rm H}}\mathbf{\tilde{w}}=P_{R}$.
\end{lem}
\textcolor{black}{In the following, a tight upper bound on the sum-rate
is derived first and three achievable sum-rate maximization relay
beamforming algorithms will be proposed. }

\subsection{Upper Bound}

A tight upper bound on the sum-rate for this CDR system is derived
in this section. An upper bound on the sum-rate for the two-way multi-antenna
AF relay system with single-antenna UEs is given in \cite{Zhang:Two_way}.
Following \cite{Zhang:Two_way}, we consider the artificial case where
the relay could use a beamforming matrix $\mathbf{W}_{1}$ optimized
for transmission to the relayed UE and a different beamforming matrix
$\mathbf{W}_{2}$ optimized for transmission to the direct UE. In
reality, we have a broadcast transmission and the same beamforming
matrix is used for both transmissions. \textcolor{black}{From the
Appendix, we know that it is optimal for the relay to transmit at
full power.} An upper bound on the sum-rate is

\begin{align}
\mathrm{\text{\ensuremath{\max_{\mathbf{W}_{1},\mathbf{W}_{2}}}}} & \:\frac{1}{2}\log_{2}\left[1+{\rm SNR_{1}}(\mathbf{W}_{1})\right]+\frac{1}{2}\log_{2}\left[1+{\rm SINR_{2}(\mathbf{W}_{2})}\right]\nonumber \\
s.t. & \;\; P(\mathbf{h}_{R1}^{{\rm H}}\mathbf{W}_{1}^{{\rm H}}\mathbf{W}_{1}\mathbf{h}_{R1}+\mathbf{h}_{RB}^{{\rm H}}\mathbf{W}_{2}^{{\rm H}}\mathbf{W}_{2}\mathbf{h}_{RB})\nonumber \\
 & \;\;\:+\kappa_{1}||\mathbf{W}_{1}||_{F}^{2}+\kappa_{2}||\mathbf{W}_{2}||_{F}^{2}=P_{R}\label{eq:upper-bound}
\end{align}
where ${\rm SNR_{1}}(\mathbf{W}_{1})$ is a function of $\mathbf{W}_{1}$
and ${\rm SINR_{2}(\mathbf{W}_{2})}$ is a function of $\mathbf{W}_{2}$.
$\kappa_{1}$ and $\kappa_{2}$ are non-negative and fulfilling $\kappa_{1}+\kappa_{2}=1$%
\footnote{If we additionally impose the constraint that $\mathbf{W}_{1}=\mathbf{W}_{2}=\mathbf{W}$,
the solution $\mathbf{W}$ will give the exact maximal sum-rate of
the CDR system.%
}. For the two different beamformers $\mathbf{W}_{1}$ and $\mathbf{W}_{2}$,
$\kappa_{1}||\mathbf{W}_{1}||_{F}^{2}$ and $\kappa_{2}||\mathbf{W}_{2}||_{F}^{2}$
represent the corresponding two fractions of power related to the
noise enhancement in the AF relaying, respectively. Denote $R(\kappa_{1},\kappa_{2})$
to be the solution to (\ref{eq:upper-bound}). This upper bound can
be tightened by minimizing $R(\kappa_{1},\kappa_{2})$ over all feasible
values of $\kappa_{1}$ and $\kappa_{2}$. When $\kappa_{1}$ and
$\kappa_{2}$ are given, (\ref{eq:upper-bound}) can be equivalently
decomposed into two independent sub problems in (\ref{eq:upper_bound_1})
where $P_{1}$ and $P_{2}$ are the total relay power consumptions
of the beamformers $\mathbf{W}_{1}$ and $\mathbf{W}_{2}$, respectively.
Therefore, $P_{1}+P_{2}=P_{R}$. $R(\kappa_{1},\kappa_{2})$ is then
derived via $R_{1}(\kappa_{1},P_{1})+R_{2}(\kappa_{2},P_{2})$ maximization
over all the feasible pairs of $P_{1}$ and $P_{2}$.

\begin{align}
R_{1}(\kappa_{1},P_{1}) & =\max_{\mathbf{W}_{1}}\,\frac{1}{2}\log_{2}\left[1+{\rm SNR_{1}}(\mathbf{W}_{1})\right]\nonumber \\
s.t. & \quad P\mathbf{h}_{R1}^{{\rm H}}\mathbf{W}_{1}^{{\rm H}}\mathbf{W}_{1}\mathbf{h}_{R1}+\kappa_{1}||\mathbf{W}_{1}||_{F}^{2}\leq P_{1}\nonumber \\
R_{2}(\kappa_{2},P_{2}) & =\max_{\mathbf{W}_{2}}\,\frac{1}{2}\log_{2}\left[1+{\rm SINR_{2}}(\mathbf{W}_{2})\right]\label{eq:upper_bound_1}\\
s.t. & \quad P\mathbf{h}_{RB}^{{\rm H}}\mathbf{W}_{2}^{{\rm H}}\mathbf{W}_{2}\mathbf{h}_{RB}+\kappa_{2}||\mathbf{W}_{2}||_{F}^{2}\leq P_{2}.\nonumber
\end{align}
The tightest upper bound is $R_{{\rm UB}}$
\[
R_{{\rm UB}}=\min_{\kappa_{1}+\kappa_{2}=1}\max_{P_{1}+P_{2}=P_{R}}R_{1}(\kappa_{1},P_{1})+R_{2}(\kappa_{2},P_{2}).
\]
According to the derivations in the Appendix, the solutions to the
two sub-problems can be derived via the generalized Rayleigh quotient.
However, no closed form solution exists for $R_{{\rm UB}}$ and numerical
search over $\kappa_{1}$, $\kappa_{2}$, $P_{1}$ and $P_{2}$ is
required. This upper bound on the sum-rate will be used to characterize
the loss resulting from the use of suboptimal optimization methods
discussed in the following.

\subsection{Beamforming Optimization Methods}

Since the achievable sum-rate maximization problem (\ref{eq:lemma-1})
is a non-convex problem \cite{Boydbook}, where global optimum solution
is difficult to obtain within reasonable computation time. This optimization
problem has generally no closed form solution. Well-known iterative
methods can be applied such as simulated annealing, genetic and branch-and-bound%
\footnote{The branch-and-bound method \cite{Boyd:BB,two-way:BB,Codreanu:BB}
can be applied to solve problem (\ref{eq:lemma-1}) and obtain the
global optimal solution. It will be treated in a future work to further
evaluate the proposed algorithms.%
} algorithms which require very high computational load. We focus on
low-complexity algorithms to avoid prohibitively high computational
complexity. The proposals will be demonstrated via simulations in
Section \textcolor{black}{\ref{Sec:sim}} to be near-optimal solutions.

\subsubsection{Adaptive Subspace Averaging Algorithm (ASS)}

We first propose a suboptimal solution based on the subspace averaging
approach \cite{Karasalo:SSA,Wong:CRB_SSA}. The concept of subspace
averaging was introduced first in a covariance matrix suboptimal estimation
with a fixed number of dominating eigenvalues.

We use a simple but loose upper bound to form our design. The cost
function in (\ref{eq:lemma-1}) can be first approximated via using
the inequality of arithmetic and geometric means

\vspace{-10pt}

\[
\frac{\tilde{\mathbf{w}}^{{\rm H}}\mathbf{A}\tilde{\mathbf{w}}}{\mathbf{\tilde{w}}^{{\rm H}}\mathbf{B}\mathbf{\tilde{w}}}\times\frac{\tilde{\mathbf{w}}^{{\rm H}}\mathbf{C}\tilde{\mathbf{w}}}{\mathbf{\tilde{w}}^{{\rm H}}\mathbf{D}\mathbf{\tilde{w}}}\leq\frac{\left(\frac{\tilde{\mathbf{w}}^{{\rm H}}\mathbf{A}\tilde{\mathbf{w}}}{\mathbf{\tilde{w}}^{{\rm H}}\mathbf{B}\mathbf{\tilde{w}}}+\frac{\tilde{\mathbf{w}}^{{\rm H}}\mathbf{C}\tilde{\mathbf{w}}}{\mathbf{\tilde{w}}^{{\rm H}}\mathbf{D}\mathbf{\tilde{w}}}\right)^{2}}{2}.
\]
An \textcolor{black}{adaptive} real value $\alpha\,(0\leq\alpha\leq1)$
is then introduced to form the averaging adaptation with respect to
$\alpha$, which will not change the optimization of the cost function
in (\ref{eq:lemma-1}). We will obtain

\vspace{-10pt}

\[
\alpha\frac{\tilde{\mathbf{w}}^{{\rm H}}\mathbf{A}\tilde{\mathbf{w}}}{\mathbf{\tilde{w}}^{{\rm H}}\mathbf{B}\mathbf{\tilde{w}}}\times(1-\alpha)\frac{\tilde{\mathbf{w}}^{{\rm H}}\mathbf{C}\tilde{\mathbf{w}}}{\mathbf{\tilde{w}}^{{\rm H}}\mathbf{D}\mathbf{\tilde{w}}}\leq\frac{\left[\alpha\frac{\tilde{\mathbf{w}}^{{\rm H}}\mathbf{A}\tilde{\mathbf{w}}}{\mathbf{\tilde{w}}^{{\rm H}}\mathbf{B}\mathbf{\tilde{w}}}+(1-\alpha)\frac{\tilde{\mathbf{w}}^{{\rm H}}\mathbf{C}\tilde{\mathbf{w}}}{\mathbf{\tilde{w}}^{{\rm H}}\mathbf{D}\mathbf{\tilde{w}}}\right]^{2}}{2}
\]
We denote $g_{1}(\tilde{\mathbf{w}})=\frac{\tilde{\mathbf{w}}^{{\rm H}}\mathbf{A}\tilde{\mathbf{w}}}{\mathbf{\tilde{w}}^{{\rm H}}\mathbf{B}\mathbf{\tilde{w}}}$
and $g_{2}(\tilde{\mathbf{w}})=\frac{\tilde{\mathbf{w}}^{{\rm H}}\mathbf{C}\tilde{\mathbf{w}}}{\mathbf{\tilde{w}}^{{\rm H}}\mathbf{D}\mathbf{\tilde{w}}}$.
We denote $\arg\max_{\tilde{\mathbf{w}}}\left[\alpha g_{1}(\tilde{\mathbf{w}})+(1-\alpha)g_{2}(\tilde{\mathbf{w}})\right]$
to be the approximated objective function for the adaptive subspace
averaging (ASA) algorithm. The two terms in the approximated objective
can be rewritten as

\vspace{-10pt}
\begin{align}
g_{1}(\tilde{\mathbf{w}}) & =\frac{\tilde{\mathbf{u}}^{{\rm H}}\mathbf{B}^{-\frac{1}{2}}\mathbf{A}\mathbf{B}^{-\frac{1}{2}}\tilde{\mathbf{u}}}{\mathbf{\tilde{u}}^{{\rm H}}\mathbf{\tilde{u}}},\;\quad\tilde{\mathbf{u}}=\mathbf{B}^{\frac{1}{2}}\tilde{\mathbf{w}}\nonumber \\
g_{2}(\tilde{\mathbf{w}}) & =\frac{\tilde{\mathbf{v}}^{{\rm H}}\mathbf{D}^{-\frac{1}{2}}\mathbf{C}\mathbf{D}^{-\frac{1}{2}}\tilde{\mathbf{v}}}{\mathbf{\tilde{v}}^{{\rm H}}\mathbf{\tilde{v}}},\:\quad\tilde{\mathbf{v}}=\mathbf{D}^{\frac{1}{2}}\tilde{\mathbf{w}}.\label{eq:g_projection}
\end{align}
 We notice that (\ref{eq:g_projection}) projecting the dominating
components of $g_{1}(\tilde{\mathbf{w}})$ and $g_{1}(\tilde{\mathbf{w}})$
onto the subspaces spanned by $\tilde{\mathbf{u}}$ and $\tilde{\mathbf{v}}$,
respectively. The two individual equivalent eigenvalue decompositions
to (\ref{eq:g_projection}) are

\vspace{-10pt}

\[
g_{1}(\tilde{\mathbf{w}})=\frac{\tilde{\mathbf{w}}^{{\rm H}}\mathbf{B}^{-1}\mathbf{A}\tilde{\mathbf{w}}}{\mathbf{\tilde{w}}^{{\rm H}}\mathbf{\tilde{w}}},\; g_{2}(\tilde{\mathbf{w}})=\frac{\tilde{\mathbf{w}}^{{\rm H}}\mathbf{D}^{-1}\mathbf{C}\tilde{\mathbf{w}}}{\mathbf{\tilde{w}}^{{\rm H}}\mathbf{\tilde{w}}}.
\]
Therefore, the adaptive subspace averaging of the approximated objective
function is expressed as $\mathbf{\Pi}=\alpha\mathbf{B}^{-1}\mathbf{A}+(1-\alpha)\mathbf{D}^{-1}\mathbf{C}$
where $\alpha\,(0\leq\alpha\leq1)$ is the adaptive parameter. The
introduction of the adaptive parameter is the novelty of this ASA
algorithm since the previous application in \cite{Wong:CRB_SSA} uses
a fixed $\alpha=0.5$. From the subspace averaging of the approximated
objective, the suboptimal ASA solution $\mathbf{\tilde{w}}$ can be
derived by solving the following problem:

\vspace{-10pt}

\[
\max_{\alpha}\left\{ \max_{\tilde{\mathbf{w}}}\left[\tilde{\mathbf{w}}^{{\rm H}}\mathbf{\Pi}\tilde{\mathbf{w}}\right]\right\} \; s.t.\;\tilde{\mathbf{w}}^{{\rm H}}\mathbf{\tilde{w}}=P_{R},\;0\leq\alpha\leq1
\]
where the solution is obtained via the principal eigenvector of $\alpha_{opt}\mathbf{B}^{-1}\mathbf{A}+(1-\alpha_{opt})\mathbf{D}^{-1}$
and then scaled to fulfill the power constraint $\tilde{\mathbf{w}}^{{\rm H}}\mathbf{\tilde{w}}=P_{R}$.
The optimal $\alpha_{opt}$ is obtained via a grid search followed
by the Nelder-Mead method%
\footnote{In Matlab, the Nelder-Mead method is implemented via the ``fminsearch''
function.%
} \cite{Jeffrey:NelderMead}. Although this algorithm relies on an
approximated objective function of the cost function in (\ref{eq:lemma-1}),
it will be shown by simulations to provide sum-rate results approaching
the upper bound.

\subsubsection{Power Iteration Algorithm (PIA)}

The second algorithm attempts to obtain a solution to the Karush-Kuhn-Tucker
(KKT) conditions. The first order necessary condition $\frac{\partial G(\tilde{\mathbf{w}})}{\partial\tilde{\mathbf{w}}}=0$
leads to $$G(\tilde{\mathbf{w}})\left[\left(\tilde{\mathbf{w}}^{{\rm H}}\mathbf{B}\tilde{\mathbf{w}}\right)\mathbf{D}+\left(\tilde{\mathbf{w}}^{{\rm H}}\mathbf{D}\tilde{\mathbf{w}}\right)\mathbf{B}\right]\mathbf{\tilde{w}}=\left[\left(\tilde{\mathbf{w}}^{{\rm H}}\mathbf{C}\tilde{\mathbf{w}}\right)\mathbf{A}+\left(\tilde{\mathbf{w}}^{{\rm H}}\mathbf{A}\tilde{\mathbf{w}}\right)\mathbf{C}\right]\mathbf{\tilde{w}}$$
which can be rewritten as $G(\tilde{\mathbf{w}})\mathbf{V}(\tilde{\mathbf{w}})\mathbf{\tilde{w}}=\mathbf{R}(\tilde{\mathbf{w}})\mathbf{\tilde{w}}$.
Notice $\mathbf{V}(\tilde{\mathbf{w}})$ and $\mathbf{R}(\tilde{\mathbf{w}})$
depend on the unknown $\tilde{\mathbf{w}}$. If the dependence could
be removed, then the optimizer $\tilde{\mathbf{w}}$ is obviously
the eigenvector corresponding to the largest eigenvalue of the matrix
$\mathbf{V}^{-1}\mathbf{R}$. However, eigenvalue decomposition of
the matrix $\left[\mathbf{V}(\tilde{\mathbf{w}})\right]^{-1}\mathbf{R}(\tilde{\mathbf{w}})$
can not be accomplished in closed form. Consequently, we propose a
power iteration algorithm (PIA) which finds the principal eigenvector
corresponding to the maximum eigenvalue in $\left[\mathbf{V}(\tilde{\mathbf{w}})\right]^{-1}\mathbf{R}(\tilde{\mathbf{w}})$
iteratively. This algorithm comes from the power iteration idea in
\cite{Lee:SR_Two_way,MatrixBook}. This proposed iterative algorithm
is described in Algorithm \ref{Flo:GPI-1}. Then the beamforming solution
is scaled to meet the power constraint $\tilde{\mathbf{w}}^{{\rm H}}\mathbf{\tilde{w}}=P_{R}$.

Since the optimization problem is non-convex, the proposed algorithm
cannot guarantee convergence. Extensive simulations have demonstrated
the convergence property: 20 iterations appear to be sufficient. In
addition, PIA provides a sub-optimal solution giving near-optimal
sum-rate results which as shown in Section \ref{Sec:sim}.

\begin{algorithm}[tbh]
\caption{Power Iteration Algorithm (PIA)}

Initialization: set $n=0$ and $\tilde{\mathbf{w}}^{{\rm (0)}}=\tilde{\mathbf{w}}^{{\rm (init)}}$

iterate

$\quad$update $n=n+1$
\begin{enumerate}
\item $\mathbf{q}^{(n)}=\left[\mathbf{V}\left(\tilde{\mathbf{w}}^{(n)}\right)\right]^{-1}$
$\times\left[\mathbf{R}\left(\tilde{\mathbf{w}}^{(n)}\right)\right]\tilde{\mathbf{w}}^{(n)}$
\item $\tilde{\mathbf{w}}^{(n+1)}=\sqrt{P_{R}}\,\mathbf{q}^{(n)}/||\mathbf{q}^{(n)}||_{2}$
\end{enumerate}
until $G(\tilde{\mathbf{w}}^{(n+1)})$ or sum-rate convergence

\label{Flo:GPI-1}
\end{algorithm}

\vspace{-10pt}

\subsubsection{Linear Space Spanning Algorithm (LSS)}

We know from Lemma 1 that solving the sum-rate maximizating problem
is equivalent to maximizaing $g(\tilde{\mathbf{w}})=g_{1}(\tilde{\mathbf{w}})g_{2}(\tilde{\mathbf{w}})$
jointly. The two beamforming vectors $\tilde{\mathbf{w}}_{1}$ and
$\tilde{\mathbf{w}}_{2}$ maximizing $g_{1}(\tilde{\mathbf{w}}_{1})$
and $g_{2}(\tilde{\mathbf{w}}_{2})$ separately could be straightforwardly
obtained.

The third low-complexity suboptimal solution is proposed based on
this observation to optimally combine the two vectors. The solution
is chosen to lie in the linear space spanned by $\tilde{\mathbf{w}}_{1}$
and $\tilde{\mathbf{w}}_{2}$,$\tilde{\mathbf{w}}_{LSS}=a\tilde{\mathbf{w}}_{1}+b\tilde{\mathbf{w}}_{2}$
where $a$ and $b$ are real value parameters. This algorithm is termed
to be the linear space spanning (LSS) algorithm. It is obvious to
see that any scaling of $a$ does not change the $\left[g_{1}(\tilde{\mathbf{w}}_{LSS})\:\, g_{2}(\tilde{\mathbf{w}}_{LSS})\right]$
maximization. Therefore, $\tilde{\mathbf{w}}_{LSS}$ is further simplified
by letting $a=1$ and $\tilde{\mathbf{w}}_{LSS}=\tilde{\mathbf{w}}_{1}+b\tilde{\mathbf{w}}_{2}$.
It is worth pointing out that the sum-rate maximization problem is
transformed into a maximization of a scalar-valued nonlinear function
$g(b)$ over one real parameter without constraints. Simulations show
there is a global maximal for $g(b)$. Again, a grid search using
the Nelder-Mead method is applied to efficiently solve the problem.
The obtained beamforming solution should be scaled to satisfy the
relay power constraint in the end.

\subsubsection{Computational Complexity}

In the adaptive subspace averaging and linear space spanning algorithms,
the optimization is over one real parameter only, the computational
complexity is lower than e.g. the branch-and-bound algorithm. In the
power iteration algorithm, the fast convergence behavior guarantees
relatively low computational complexity.

\textcolor{red}{}

\section{Numerical Results\label{Sec:sim} }

In this section, we present simulation results for the sum-rate. We
assume the relay and the BS have the same transmit power, i.e. $P_{R}=P$.
The relay beamforming designs targeting either $\text{SNR}_{1}$ maximization
or $\text{SINR}_{2}$ maximization are also included. In addition,
to assess the effect of linear relay beamforming, the trivial pure
amplification relaying $\mathbf{W}=\sqrt{P_{R}/\left(P\mathbf{h}_{RB}^{{\rm H}}\mathbf{h}_{RB}+P\mathbf{h}_{R1}^{{\rm H}}\mathbf{h}_{R1}+||\mathbb{I}||_{F}^{2}\right)}\mathbb{I}$
is also considered. The benchmark with single antenna relay is included
to evaluate the gain from using multiple relay antennas.

We compare the sum-rate performance with respect to different
relay antenna numbers. The proposed algorithms are better than the
relay beamforming design targeting either $\text{SNR}_{1}$ or $\text{SINR}_{2}$
maximization and performing close to the upper bound. The performance
gap between the proposals and the upper bound becomes smaller when
the number of relay antenna is large. Moreover, PIA performs the closest
to the tight upper bound on the sum-rate among all the three proposals
and ASA has a tiny performance loss compared to PIA. Therefore, PIA
is an efficient tool to address sum-rate maximization of the multi-antenna
AF CDR system, although it is sub-optimal. It is also observed that
the pure amplification relaying causes a significant performance loss.
The sum-rate gain from the multiple-antenna relay beamforming is obvious,
compared to the single antenna relay transmission. With the increase
of the number of antennas at relay, we can see a clear increase in
the sum-rate performance.

%\begin{figure}[tbh]
%\vspace{-8pt}
%
%\centering \includegraphics[scale=0.55]{SR_M2}
%
%\caption{Sum-rate performance ($M=2$).}
%
%
%\vspace{-10pt}
%
%\label{Flo:Sum_rate-M=00003D2}
%\end{figure}
%
%
%\begin{figure}[tbh]
%\vspace{-10pt}
%
%\centering \includegraphics[scale=0.55]{SR_M4}
%
%\caption{Sum-rate performance ($M=4$).}
%
%
%\vspace{-16pt}
%
%\label{Flo:Sum_rate-M=00003D4}
%\end{figure}

\begin{figure*}
\begin{align}
\arg\max_{\mathbf{W}} & \:\left(1+\frac{P\left\Vert \mathbf{h}_{BR}\mathbf{W}\mathbf{h}_{R1}\right\Vert _{2}^{2}}{\mathbf{h}_{BR}\mathbf{W}\mathbf{W}^{{\rm H}}\mathbf{h}_{BR}^{{\rm H}}+1}\right)\times\left(1+\frac{P\left\Vert h_{2B}\mathbf{h}_{2R}\mathbf{W}\mathbf{h}_{R1}-h_{21}\mathbf{h}_{2R}\mathbf{W}\mathbf{h}_{RB}\right\Vert _{2}^{2}}{|h_{21}|^{2}\left(\mathbf{h}_{2R}\mathbf{W}\mathbf{W}^{{\rm H}}\mathbf{h}_{2R}^{{\rm H}}+1\right)+\left\Vert \mathbf{h}_{2R}\mathbf{W}\mathbf{h}_{R1}\right\Vert _{2}^{2}}\right)\nonumber \\
s.t. & \quad P(\mathbf{h}_{RB}^{{\rm H}}\mathbf{W}^{{\rm H}}\mathbf{W}\mathbf{h}_{RB}+\mathbf{h}_{R1}^{{\rm H}}\mathbf{W}^{{\rm H}}\mathbf{W}\mathbf{h}_{R1})+||\mathbf{W}||_{F}^{2}=P_{R}\label{eq:sum_rate_max_criterion_reform}
\end{align}
\end{figure*}

\begin{figure*}[!t]
\vspace{-15pt}

\begin{align}
\arg\max_{\mathbf{w}} & \:\left[1+\frac{P\mathbf{w}^{{\rm H}}(\mathbf{h}_{R1}^{{\rm T}}\otimes\mathbf{h}_{BR})^{{\rm H}}(\mathbf{h}_{R1}^{{\rm T}}\otimes\mathbf{h}_{BR})\mathbf{w}}{\mathbf{w}^{{\rm H}}(\mathbb{I}\otimes\mathbf{h}_{BR})^{{\rm H}}(\mathbb{I}\otimes\mathbf{h}_{BR})\mathbf{w}+1}\right]\times\left[1+\frac{P\mathbf{w}^{{\rm H}}\mathbf{f}\mathbf{\mathbf{f}^{{\rm H}}}\mathbf{w}}{|h_{21}|^{2}\mathbf{w}^{{\rm H}}\mathbf{C}_{1}^{{\rm H}}\mathbf{C}_{1}\mathbf{w}+|h_{21}|^{2}+\mathbf{w}^{{\rm H}}\mathbf{a}\mathbf{a}^{{\rm H}}\mathbf{w}}\right]\nonumber \\
s.t. & \quad\mathbf{w}^{{\rm H}}\left[P(\mathbf{h}_{RB}^{{\rm T}}\otimes\mathbb{I})^{{\rm H}}(\mathbf{h}_{RB}^{{\rm T}}\otimes\mathbb{I})+P(\mathbf{h}_{R1}^{{\rm T}}\otimes\mathbb{I})^{{\rm H}}(\mathbf{h}_{R1}^{{\rm T}}\otimes\mathbb{I})\mathbf{+\mathbb{I}}\right]\mathbf{w}=P_{R}\label{eq:sum_rate_max_criterion_reform-1}\\
 & \quad\mathbf{a}=(\mathbf{h}_{R1}^{{\rm T}}\otimes\mathbf{h}_{2R})^{{\rm H}},\quad\mathbf{C}_{1}=\mathbb{I}\otimes\mathbf{h}_{2R},\quad\mathbf{f}^{{\rm H}}=h_{2B}(\mathbf{h}_{R1}^{{\rm T}}\otimes\mathbf{h}_{2R})-h_{21}(\mathbf{h}_{RB}^{{\rm T}}\otimes\mathbf{h}_{2R})\nonumber
\end{align}

\vspace{-15pt}

\begin{align}
\arg\max_{\tilde{\mathbf{w}}} & \:\frac{\tilde{\mathbf{w}}^{{\rm H}}\left\{ \mathbf{\mathbf{J}}^{{\rm -H}}\left[P(\mathbf{h}_{R1}^{{\rm T}}\otimes\mathbf{h}_{BR})^{{\rm H}}(\mathbf{h}_{R1}^{{\rm T}}\otimes\mathbf{h}_{BR})+(\mathbb{I}\otimes\mathbf{h}_{BR})^{{\rm H}}(\mathbb{I}\otimes\mathbf{h}_{BR})\right]\mathbf{\mathbf{J}}^{{\rm -1}}+\frac{1}{P_{R}}\mathbb{I}\right\} \tilde{\mathbf{w}}}{\mathbf{\tilde{w}}^{{\rm H}}\left[\mathbf{\mathbf{J}}^{{\rm -H}}(\mathbb{I}\otimes\mathbf{h}_{BR})^{{\rm H}}(\mathbb{I}\otimes\mathbf{h}_{BR})\mathbf{\mathbf{J}}^{{\rm -1}}+\frac{1}{P_{R}}\mathbb{I}\right]\mathbf{\tilde{w}}}\nonumber \\
 & \;\;\times\frac{\tilde{\mathbf{w}}^{{\rm H}}\left[\mathbf{\mathbf{J}}^{{\rm -H}}\left(|h_{21}|^{2}\mathbf{C}_{1}^{{\rm H}}\mathbf{C}_{1}+\mathbf{a}\mathbf{a}^{{\rm H}}+P\mathbf{f}\mathbf{\mathbf{f}^{{\rm H}}}\right)\mathbf{\mathbf{J}}^{{\rm -1}}+\frac{|h_{21}|^{2}}{P_{R}}\mathbb{I}\right]\tilde{\mathbf{w}}}{\mathbf{\tilde{w}}^{{\rm H}}\left[\mathbf{\mathbf{J}}^{{\rm -H}}\left(|h_{21}|^{2}\mathbf{C}_{1}^{{\rm H}}\mathbf{C}_{1}+\mathbf{a}\mathbf{a}^{{\rm H}}\right)\mathbf{\mathbf{J}}^{{\rm -1}}+\frac{|h_{21}|^{2}}{P_{R}}\mathbb{I}\right]\mathbf{\tilde{w}}}\label{eq:sum_rate_max_criterion_reform-2}\\
s.t. & \quad\tilde{\mathbf{w}}^{{\rm H}}\mathbf{\tilde{w}}=P_{R}\nonumber
\end{align}

\vspace{-10pt}
\noindent\rule{\textwidth}{0.8pt}
\vspace{-30pt}
\end{figure*}

\section{Conclusions}

We focus on the relay beamforming design for sum-rate maximization
of the AF CDR system. We characterize a tight upper bound on the sum-rate
and propose three low-complexity but efficient algorithms to approach
the achievable sum-rate maximum. Numerical results confirm that the
proposals give comparable sum-rate and perform close to the tight
upper bound. PIA is identified to be the best giving near-optimal
sum-rate performance. An obvious sum-rate increase from the usage
of multiple relay antennas is also observed.

\appendices{}

\section{Proof of Lemma 1}
\begin{IEEEproof}
The problem (\ref{eq:sum_rate_max_criterion}) can be formulated in
(\ref{eq:sum_rate_max_criterion_reform}). It can be easily proved
that the relay power constraint in (\ref{eq:sum_rate_max_criterion_reform})
should be met with equality at the optimum. Therefore, it is sum-rate
optimal for the relay to transmit at full power $P_{R}$. In order
to rewrite the optimization cost function in a simple way, the beamforming
matrix $\mathbf{W}$ is converted into a vector form using the vectorization
operation, $\mathbf{w=}{\rm vec}(\mathbf{W})$. With the property
${\rm vec}(\mathbf{MWN})=(\mathbf{N}^{{\rm T}}\otimes\mathbf{M}){\rm vec}(\mathbf{W})$,
we can rewrite the problem in (\ref{eq:sum_rate_max_criterion_reform-1})
where the relay power inequality constraint is replaced by an equality
constraint. We further introduce $\mathbf{J}$ from the Cholesky decomposition
\[
P(\mathbf{h}_{RB}^{{\rm T}}\otimes\mathbb{I})^{{\rm H}}(\mathbf{h}_{RB}^{{\rm T}}\otimes\mathbb{I})+P(\mathbf{h}_{R1}^{{\rm T}}\otimes\mathbb{I})^{{\rm H}}(\mathbf{h}_{R1}^{{\rm T}}\otimes\mathbb{I})\mathbf{+\mathbb{I}\triangleq\mathbf{J}}^{{\rm H}}\mathbf{J}.
\]
We let $\mathbf{\tilde{w}=Jw}$. When applying $\mathbf{w=J^{-1}\tilde{w}}$,
the problem can be finally reformulated in (\ref{eq:sum_rate_max_criterion_reform-2}).
We further observe that the norm of $\mathbf{\tilde{w}}$ does not
influence the maximization at all. Hence, the constraint can be ignored.
This transforms the problem (\ref{eq:sum_rate_max_criterion_reform-2})
into an unconstrained maximization problem. After some mathematical
manipulations, it can be readily observed the reformulated sum-rate
maximization beamforming design problem is in the form of $\arg\max_{\tilde{\mathbf{w}}}G(\tilde{\mathbf{w}})=\arg\max_{\tilde{\mathbf{w}}}\left[\frac{\tilde{\mathbf{w}}^{{\rm H}}\mathbf{A}\tilde{\mathbf{w}}}{\mathbf{\tilde{w}}^{{\rm H}}\mathbf{B}\mathbf{\tilde{w}}}\times\frac{\tilde{\mathbf{w}}^{{\rm H}}\mathbf{C}\tilde{\mathbf{w}}}{\mathbf{\tilde{w}}^{{\rm H}}\mathbf{D}\mathbf{\tilde{w}}}\right]$
where matrices $\mathbf{A}$, $\mathbf{B}$, $\mathbf{C}$, and $\mathbf{D}$
are not dependent on $\tilde{\mathbf{w}}$. We scale $\mathbf{\tilde{w}}$
in the end to fulfill the relay power constraint. This completes the
proof.
\end{IEEEproof}

\section*{}

\bibliographystyle{IEEEtran} \phantomsection\addcontentsline{toc}{section}{\refname}
\bibliographystyle{IEEEtran}
\bibliography{IEEEabrv,IEEEexample,capacity}

\end{document}